\documentclass[12pt,ta4paper]{article}

\usepackage{siunitx}
\usepackage[colorlinks,bookmarksopen,bookmarksnumbered,citecolor=red,urlcolor=red]{hyperref}
\usepackage{amsmath}
\usepackage{textcomp} % this enables \textdegree
\usepackage{setspace} % for peer-review
\usepackage{natbib}
\usepackage{graphicx}

\usepackage{a4wide}
\usepackage[T1]{fontenc}
\usepackage[utf8]{inputenc}
\usepackage{tgschola}

\title{
Elimination of spectral blocking by ensuring rotation-free property of
discretised pressure gradient within a spectral semi-implicit semi-Lagrangian
global atmospheric model
}

\author{Masashi Ujiie\thanks{Numerical Prediction Division, Japan Meteorological Agency, Tokyo, Japan} 
  \and  Daisuke Hotta\thanks{Meteorological Research Institute, Japan Meteorological Agency, Tsukuba, Japan}}

\begin{document}
\singlespacing
\maketitle

{\it
This is the submitted version of the following article: Ujiie and Hotta (2019; {\it QJRMS}), which has been published in final  form at \verb|https://doi.org/10.1002/qj.3636|. This article may be used for non-commercial purposes in  accordance with Wiley Terms and Conditions for Self-Archiving.
}

\begin{abstract}
The widely-adopted discretisation of the horizontal pressure gradient
term formulated by Simmons and Burridge (1981) for atmospheric models on
$\sigma$-$p$ hybrid vertical coordinate is found to incur spectral
blocking for rotational wind components at high vertical levels when
used in a spectral semi-Lagrangian model run on a linear grid.
% The spectral blocking becomes
% particularly severe when the resolution is high and when used with
% realistically steep orography. 
A remedy to this issue is proposed and tested using a spectral
semi-implicit semi-Lagrangian hydrostatic primitive equations model. The
proposed method removes aliasing errors at high wavenumbers by ensuring
that the rotation-free property of the pressure gradient term on isobaric
surface, a feature possessed by the continuous system, is preserved in
the discretised system, which highlights the significance of mimetic
discretisation within the context of numerical weather prediction
models.
% Please include a maximum of seven keywords
%\keywords{spectral method; spectral blocking; pressure gradient; mimetic discretisation}
\end{abstract}

%%%%% commands for editting, should be disabled before submission %%%%
%\let\origlabel\label
%\renewcommand{\label}[1]{\ \mbox{\tt [#1]} \origlabel{#1}  }

%\linenumbers

%% some command definitions
\newcommand{\ie}{\textit{i.e.}}

%\doublespacing

\section{Introduction}

``Spectral blocking'' is a phenomenon often encountered in numerical
time-marching solution of nonlinear partial differential equations that
is characterised by a turn-up of power-spectra near the truncation limit
\citep{Boyd}. On a map, spectral blocking often manifests itself as
small-scale noises which, if left uncontrolled, can destabilise model
integration. Spectral blocking is also undesirable because it means that
the numerical solution in the high wavenumber range is dominated by
noises rather than physically meaningful signals
\citep[e.g.][]{LanderHoskins97}. It is therefore of prime importance to
identify the cause of the spectral blocking and to take appropriate
measures to eliminate or at least mitigate it.  Note that, unlike its
name may suggest, spectral blocking can occur with any discretisation
method, not limited to the spectral method.

Spectral blocking has been recognised early in the history of climate
and numerical weather prediction (NWP) model development when the
primary source of nonlinearity was the quadratic Eulerian advection
term, and various methods to counteract it have been proposed. For a
finite-difference model, \cite{Arakawa66} devised a Jacobian (advection)
discretisation that ensures energy and enstrophy conservation, whereby
preventing small-scale noises to grow indefinitely; for a spectral
model, \cite{Orszag70} introduced quadratic grid truncation rule which
eliminates aliasing from quadratic terms by throwing away the highest
one-third of the spectra that can be represented with a given model
grid. Following \cite{Orszag70}, all spectral atmospheric models with
Eulerian advection scheme adopted quadratic (or higher-order) grid
truncation. Later, when semi-Lagrangian advection schemes
\citep{Robert81,ritchie88} were introduced that do not explicitly treat
the quadratic advection terms, these schemes were shown to be able to
control aliasing even with linear grid truncation
\citep{CoteStaniforth88}, which led many operational spectral
semi-Lagrangian models, including European Centre for Medium-range
Weather Forecasts (ECMWF)'s Integrated Forecasting System (IFS),
National Centers for Environmental Prediction (NCEP)'s Global
Forecasting System (GFS) and Japan Meteorological Agency (JMA)'s Global
Spectral Model (GSM), to adopt linear grids
\citep{Hortal02,Katayama05,Sela10}. We remark here that the advection
terms, while being the dominant ones, are only one of the many sources
of nonlinearity in the atmospheric governing equations (the others
include physical parameterisation, pressure gradient terms in the
momentum equations, and the adiabatic heating term in the thermodynamic
equation, just to name a few), so that the absence of advection terms in
the semi-Lagrangian schemes does not necessarily guarantee absence of
the aliasing problem.

In JMA-GSM, the presence of spectral blocking became increasingly
apparent as the resolution increased. As an example, Left panels of
Figure \ref{fig:spectra} show the rotational and divergent component of
the kinetic energy spectra computed for GSM's two-day forecast
initialised at a particular date. Spectral blocking appears for the
rotational component (thick lines) and curiously, it is stronger in the
upper levels (Figure \ref{fig:spectra}a) than in the lower levels
(Figure \ref{fig:spectra}b). After an extensive investigation at JMA
aiming at understanding this issue, the pressure gradient discretisation
in the vertical, formulated following \citet[SB81, hereafter]{SB81},
was found to be primarily responsible for this spectral blocking.

In this note, we describe why the discretisation of the pressure
gradient terms adopted in JMA-GSM following SB81 can produce spectral
blocking in the rotational wind component, and show how this problem can
be remedied. As we show later in this note, the key is to ensure that
the vector calculus identity (that the curl of gradient of any scalar
field is zero) is preserved in the discretised system, which highlights
the importance of ``mimetic discretisation'' in the context of NWP and
climate modelling.

The rest of this note is structured as follows. Section \ref{sec:SB81}
reviews the discretisation of the pressure gradient terms on
$\sigma$-$p$ hybrid coordinate given in SB81 and discusses how,
depending on specific implementation, it may cause spectral blocking
when combined with a spectral horizontal discretisation. Section
\ref{sec:remedy} proposes a method to alleviate this issue. Sections
\ref{sec:idealised-exp} and \ref{sec:nwp-exp} describe the setup of
idealised and realistic experiments along with the results. Section
\ref{sec:conclusions} concludes the note with discussions on its
implication for future development of dynamical cores.
%
% One may wonder how we narrowed down to pressure gradient out of many
% possibilities. Some clues that led us to focus on pressure gradient are
% summarised in Appendix.

The pressure gradient formulation adopted in JMA-GSM is typical of many
spectral semi-implicit semi-Lagrangian (SISL) atmospheric models,
including ECMWF-IFS \citep{IFSdoc}, SISL version of National Center for
Atmospheric Research (NCAR) Community Atmospheric Model (CAM)
\citep{WO94}, M\'{e}t\'{e}o France's ALADIN regional model
\citep{Benard10}, and NCEP-GFS \citep{Sela10}. We thus anticipate that
the solution we found can also be helpful to other models in resolving
(if present) similar issues.

\section{Pressure gradient discretisation on $\sigma$-$p$ hybrid vertical levels following SB81}
\label{sec:SB81}

On a terrain-following $\sigma$-$p$ hybrid vertical coordinate
parameterised by $\eta\in [0,1]$ where $\eta$ is defined such that the
pressure $p$ can be expressed as $p=A(\eta)+B(\eta)p_s$ in terms of
surface pressure $p_s$ and some functions $A$ and $B$, the hydrostatic
pressure gradient  $\mathbf{F}_{\mathrm{pgrad}}=-\nabla_p\Phi$ on
the right-hand side (RHS) of the momentum equations is expressed as
\begin{align}
\mathbf{F}_{\mathrm{pgrad}} =& -\nabla_p\Phi  = -\nabla_\eta \Phi - R_dT_v\nabla_\eta \ln{p}, \quad \mbox{with} \label{eq:pgrad-continuous}\\
 \Phi & = \Phi_s + \int_p^{p_s} R_dT_vd\ln{p}, \label{eq:hydro}
\end{align}
where $\Phi, T_v$ and $R_d$ are, respectively, the geopotential, virtual
temperature, and the gas constant for dry atmosphere. The subscript $s$
signifies values at the surface, and the subscripts $p$ and
$\eta$ given to the gradient operator $\nabla$ signify that the
derivative is taken along the isobaric and iso-$\eta$ surfaces,
respectively. Note that the pressure gradient terms become rotation-free
on upper atmosphere where the $\eta$-levels become isobaric
($B(\eta)=0$) because the first term on Equation
\ref{eq:pgrad-continuous} is rotation-free from vector calculus identity
($\nabla_\eta\times\nabla_\eta\equiv0$) and
 the second term is
identically zero for isobaric surfaces:
\begin{align}
 \nabla_\eta \times \mathbf{F}_{\mathrm{pgrad}} = \mathbf{0} \quad \mbox{for}\quad \eta\quad  \mbox{such that}\quad B(\eta)=0. \label{eq:rot-free}
\end{align}

SB81 derived a vertical discretisation on $\eta$ levels that conserves
globally averaged angular momentum. Counting the levels from the surface
($k=1$) up to the model top ($k=k_\mathrm{max}$), the hydrostatic relation
(Equation \ref{eq:hydro}) is discretised as
\begin{align}
 \Phi_k =\Phi_s + \sum_{l=1}^{k-1}R_dT_{v,l}
 \ln\left(\frac{p_{l-1/2}}{p_{l+1/2}}\right)+\alpha_kR_dT_{v,k} \label{eq:hydro-disc}
\end{align}
to give
\begin{align}
 -\nabla_\eta\Phi_k =-\nabla_\eta \Phi_s 
- \sum_{l=1}^{k-1}R_d \nabla_\eta\left[T_{v,l} \ln\left(\frac{p_{l-1/2}}{p_{l+1/2}}\right) \right]
-\alpha_kR_d\nabla_\eta T_{v,k}, \label{eq:gradphi-disc}
\end{align}
where the subscripts $k$ and $l$ denote discretised values at $k$-th and
$l$-th full levels, the subscripts $l\pm1/2$ denote the values at half
levels, and $\alpha_k$ is defined as
 $\left[1-\frac{p_{k+1/2}}{\Delta p_k}\ln\left(\frac{p_{k-1/2}}{p_{k+1/2}}\right)\right]$
 (for $k\neq k_\mathrm{max}$) where $\Delta p_k:=p_{k-1/2}-p_{k+1/2}$ and $\alpha_{k_\mathrm{max}}=\ln{2}$. 
Similarly the second term in Equation \ref{eq:pgrad-continuous} at full levels is
discretised as
\begin{align}
- \left({R_{d}T_{v}}\nabla_{\eta}\ln{p}\right)_{k} &=
- \frac{R_{d}T_{v,k}}{\Delta p_{k}}
   \left[
      \ln \frac{p_{k-1/2}}{p_{k+1/2}}
      \nabla_{\eta}p_{k+1/2}
      + \alpha_{k}  \nabla_{\eta}\Delta p_{k}
\right]. \label{eq:pgrad-second-disc}
\end{align}
The SB81 discretisation shown above is adopted by many $\eta$-coordinate
models but how the pressure gradient terms are precisely computed varies
depending on specific implementation of each model.

In spectral models, naive evaluation of Equation
\ref{eq:pgrad-second-disc} may require a spectral transform of a
three-dimensional variable $p$ (or $\ln{p}$) to compute its gradient,
which is not economical since $\nabla_\eta p$ (or $\nabla_\eta \ln{p}$)
is not used elsewhere in the model. SB81 suggested to economise
computation by using the vertical coordinate definition
\begin{align}
 p_{k-1/2} & = A_{k-1/2} + B_{k-1/2}p_s \label{eq:eta-def}
\end{align}
to express the horizontal derivatives in Equation
\ref{eq:pgrad-second-disc} in terms of $\nabla_\eta T_{v,k}$ and
$\nabla_\eta p_s$ (or $\nabla_\eta \ln{p_s}$). In semi-Lagrangian models
based on `$U$-$V$' formulation \citep{ritchie88,temperton91}, we can
further avoid spectral transform on another three-dimensional variable
$\Phi$ by applying a similar strategy on Equation \ref{eq:gradphi-disc};
symbolically, the pressure gradient terms are expressed as
\begin{align}
\mathbf{F}_{\mathrm{pgrad},k}
&= -\nabla_\eta\Phi_k - \left({R_{d}T_{v}}\nabla_{\eta}\ln{p}\right)_{k}   \label{eq:pgrad-disc-1}\\
\nabla_\eta\Phi_k  
&= \nabla_\eta\Phi_s +\sum_{l=1}^{k}F_{l,k}(p_s,T_v)\nabla_\eta T_{v,l} + G_k(p_s,T_v)\nabla_\eta\ln{p_s} \label{eq:pgrad-disc-2}\\
\left({R_{d}T_{v}}\nabla_{\eta}\ln{p}\right)_{k} 
&= \sum_{l=1}^{k}H_{l,k}(p_s,T_v)\nabla_\eta T_{v,l} + I_k(p_s,T_v)\nabla_\eta\ln{p_s}. \label{eq:pgrad-disc-3}
\end{align}
where the derivatives $\nabla_\eta \Phi_s, \nabla_\eta T_{v,k}$ and
$\nabla_\eta \ln{p_s}$ are evaluated spectrally.  Examples of precise
expressions can be found in, e.g., \cite{IFSdoc} and \cite{Sela10}. This
way, the number of variables to be represented spectrally is minimised,
hence necessitating minimal numbers of grid-to-wave and wave-to-grid
transforms per each time step.

JMA-GSM, before its May 2017 update \citep{Yonehara18}, adopted a
similar strategy but further simplified the expression by exploiting the
fact that the second term on the RHS of Equation \ref{eq:gradphi-disc}
and the second term on the RHS of Equation \ref{eq:pgrad-second-disc},
if expanded, share common terms that, when combined, cancel each
other. By cancelling them out, the expression for the pressure gradient
terms reduces to
\begin{align} \begin{split}
\mathbf{F}_{\mathrm{pgrad},k}
=
- \nabla_{\eta} \Phi_{s}
-  \sum_{l=1}^{k-1}R_{d}
    \ln\left( \frac{p_{l-1/2}}{p_{l+1/2}}\right)
    \nabla_{\eta} T_{v,l}  \\
-    \sum_{l=1}^{k-1}R_{d} T_{v,l} \left(\frac{B_{l-1/2}}{p_{l-1/2}} 
                - \frac{B_{l+1/2}}{p_{l+1/2}} \right)
    \nabla_{\eta} p_{s} \\
-
   \alpha_{k} R_{d} \nabla_{\eta} T_{v,k}
   - R_{d}T_{v,k}\frac{B_{k-1/2}}{p_{k-1/2}}\nabla_{\eta} p_{s},
  \label{eq:pgrad-gsm1603}
\end{split} \end{align}
where the derivatives $\nabla_\eta \Phi_s, \nabla_\eta T_{v,k}$ and
$\nabla_\eta p_s$ are evaluated spectrally. This pre-cancelled
formulation avoids loss of accuracy due to cancellation of significant
digits and should be particularly helpful over steep orography.

As we remarked at the beginning of this section, the pressure gradient
terms on isobaric levels are rotation-free in the continuous system
(Equation \ref{eq:rot-free}). The discrete analogue as computed from
Equations \ref{eq:pgrad-disc-1}-\ref{eq:pgrad-disc-3} or Equation
\ref{eq:pgrad-gsm1603}, however, does not preserve this property even
though the horizontal spectral discretisation guarantees the
rotation-free property of scalar gradients
($\nabla_\eta\times\nabla_\eta(\cdot)=\mathbf{0}$). Since the pressure
gradient terms are nonlinear with respect to the model's prognostic
variables, the purely numerical noises in rotational component induced
by the inability of the pressure gradient discretisation to preserve the
rotation-free property contain high-wavenumber components beyond the
truncation limit, which alias back onto the resolved high-wavenumber
spectra of the solution which then can accumulate over time steps, eventually
manifesting itself as spectral blocking. 

Note that this problem does not show up in semi-implicit `$\zeta$-$D$'
models \citep{HS75} since, in `$\zeta$-$D$' formulation, the pressure
gradient terms only appear in the divergence equation in the form of a
Laplacian. Nonlinear aliasing does occur for the divergent component but
it is well controlled by the selective high-wavenumber damping inherent
in the semi-implicit Helmholtz solver.

\section{Modified pressure gradient discretisation that alleviates spectral blocking}
\label{sec:remedy}

\begin{figure}[htbp]
 \centering
 \includegraphics[width=0.4\textwidth]{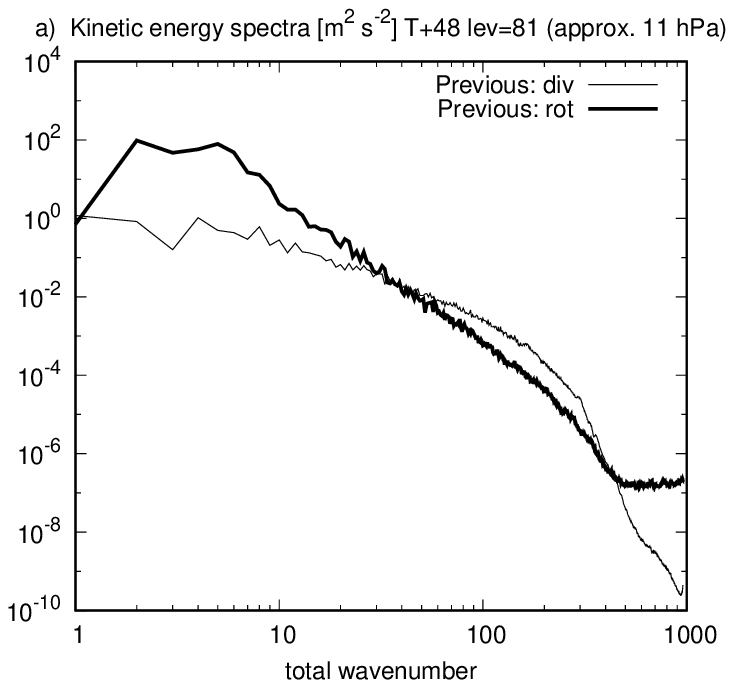}    % originally sp_GSM1603_81.eps
 \includegraphics[width=0.4\textwidth]{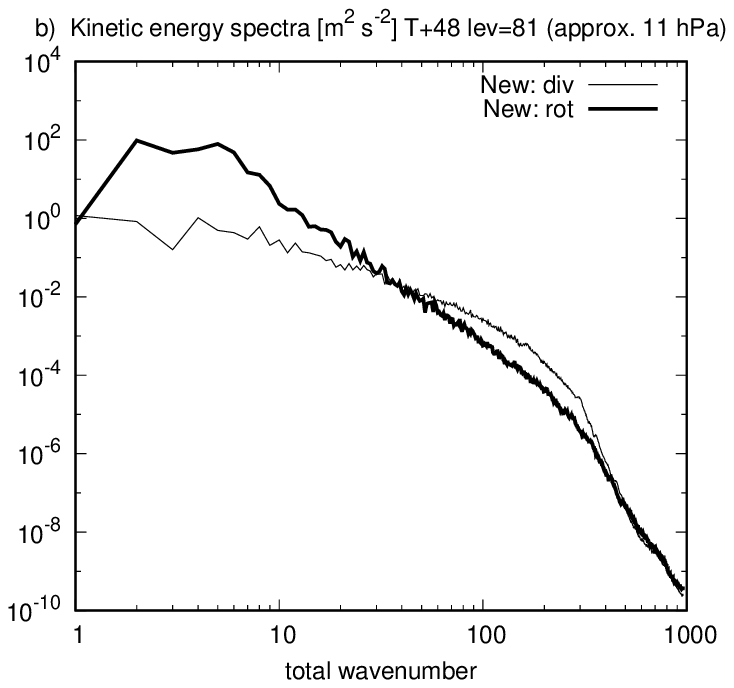} \\ % originally sp_GSM1705_81.eps
 \includegraphics[width=0.4\textwidth]{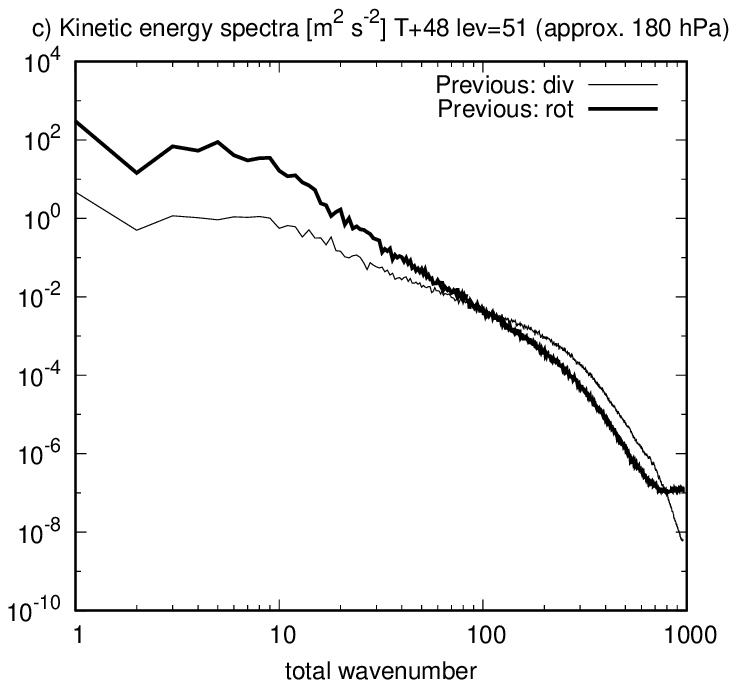}    % originally sp_GSM1603_51.eps
 \includegraphics[width=0.4\textwidth]{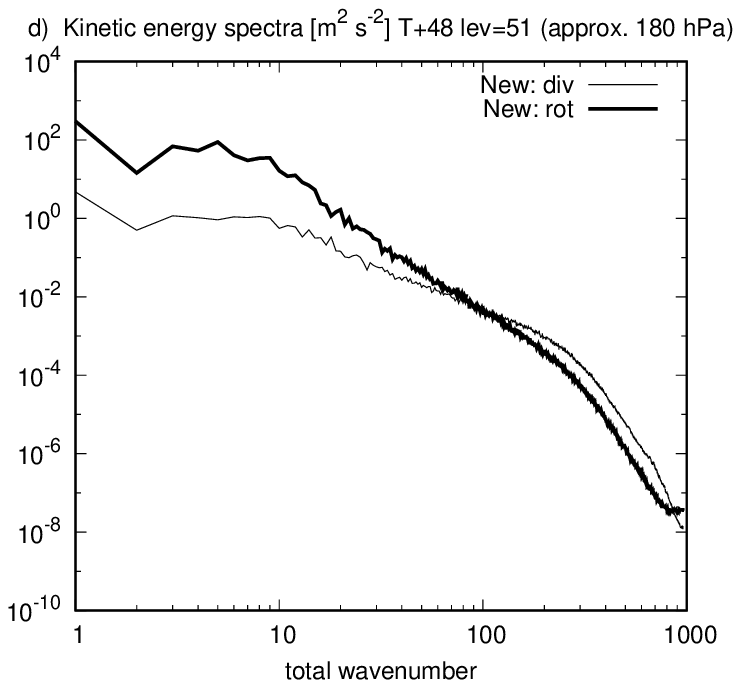}    % originally sp_GSM1705_51.eps
 \caption{
 Kinetic energy spectra of 48-hour forecasts produced by JMA-GSM with
 (a,c) previous and (b,d) new pressure gradient discretisation schemes,
 for (a,b) an upper level (81st model level, $\sim$ 11 hPa) and (c,d) a
 middle level (51st model level, $\sim$ 180 hPa). On each panel, the
 rotational and divergent components are plotted, respectively, with
 thick and thin lines. Units are $\mathrm{m}^2\mathrm{s}^{-2}$. Note the
 log scale on both axes. The forecasts are initialised with JMA's
 operational deterministic analysis valid at December 25, 2018, 12 UTC.
 }
 \label{fig:spectra}
\end{figure}

As we described in Section \ref{sec:SB81}, the failure of the pressure
gradient discretisation to preserve its rotation-free property on
isobaric levels can result in nonlinear aliasing. This finding is not
new, and \cite{Wedi13} and \cite{Wedi14} reported that a symptom similar
to ours depicted in Figure \ref{fig:spectra} was also found in
ECMWF-IFS; they identified the rotational component of the pressure
gradient terms as the primary source of aliasing, and further proposed
two solutions to this issue, one being to apply a ``de-aliasing'' filter
that effectively removes the upper one-third of the spectra of
rotational component of the pressure gradient terms, and the other being
to adopt a higher order grid truncation.
 
Here we propose an alternative solution.  The rotation-free property on
iso-baric levels (Equation \ref{eq:rot-free}) can be assured if we first
compute the full-level geopotential $\Phi_k$ with Equation
\ref{eq:hydro-disc} in grid space and then evaluate its gradient
$\nabla_\eta\Phi_k$ by spectral transform, and finally combine it with
the rest of the pressure gradient (Equation \ref{eq:pgrad-second-disc})
with $\nabla_\eta p_{k+1/2}$ and $\nabla_\eta \Delta p_{k}$ expressed in
terms of $\nabla_\eta \ln{p_s}$ using Equation \ref{eq:eta-def}. This
way, the absence of rotation on isobaric levels is automatically assured
since the spectral evaluation of the horizontal gradient guarantees the
identity $\nabla_\eta\times\nabla_\eta =\mathbf{0}$, and the second part
coming from Equation \ref{eq:pgrad-second-disc} is identically zero on
isobaric levels.

\begin{figure}[htbp]
 \centering
 \includegraphics[width=0.35\textwidth,angle=270]{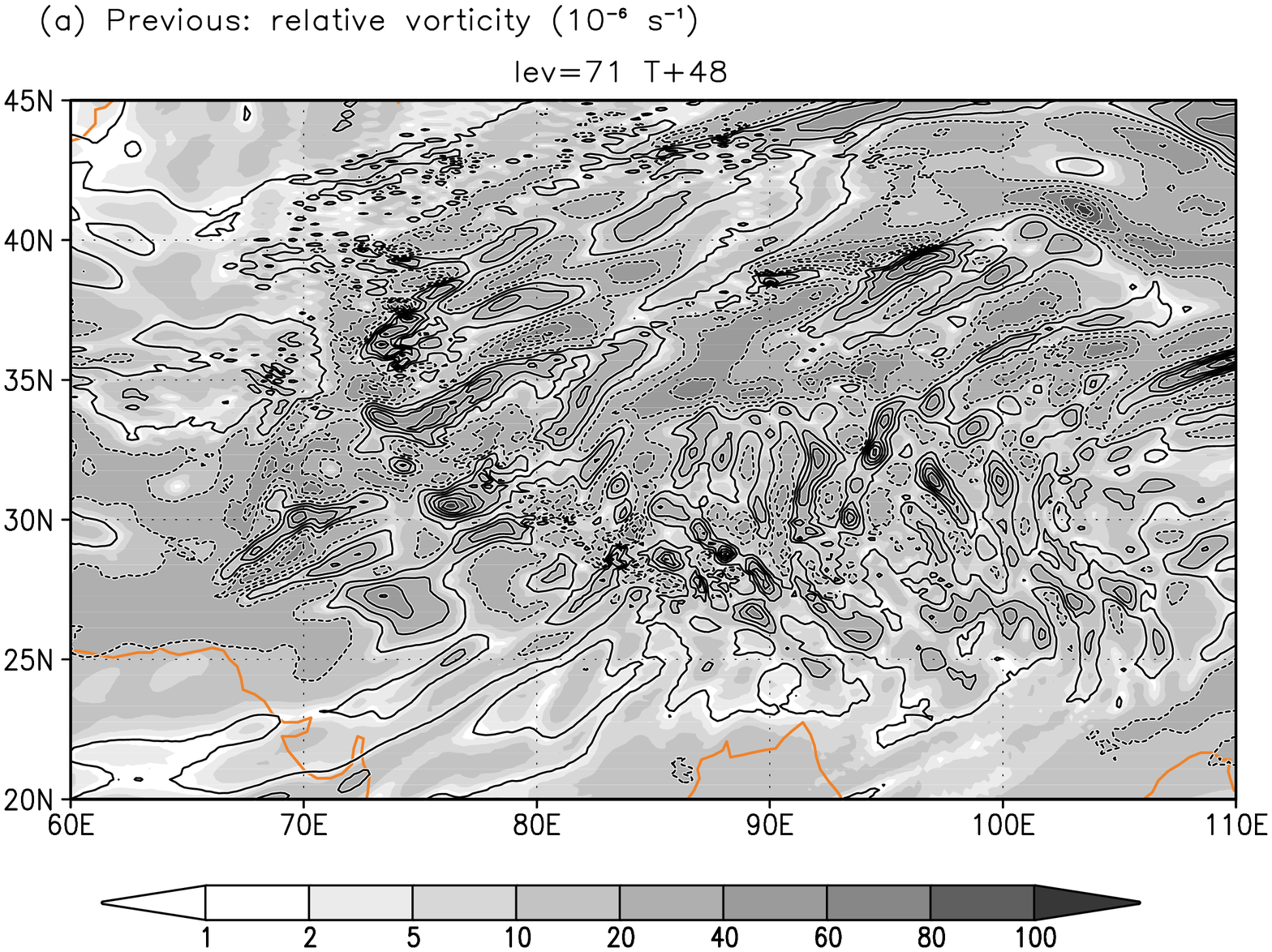}    % originally vor_GSM1603.eps
 \includegraphics[width=0.35\textwidth,angle=270]{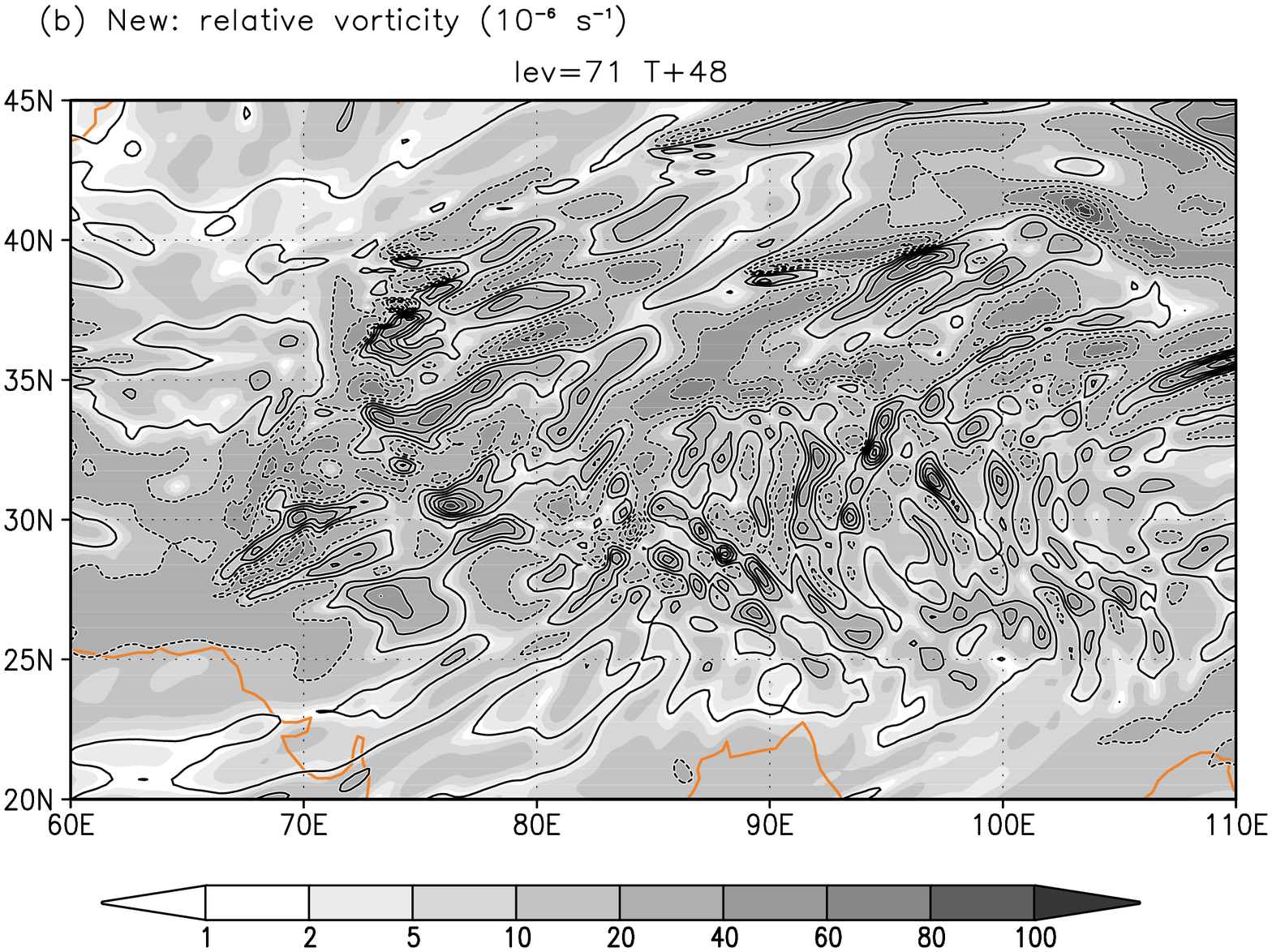} \\ % originally vor_GSM1705.eps
 \includegraphics[width=0.35\textwidth,angle=270]{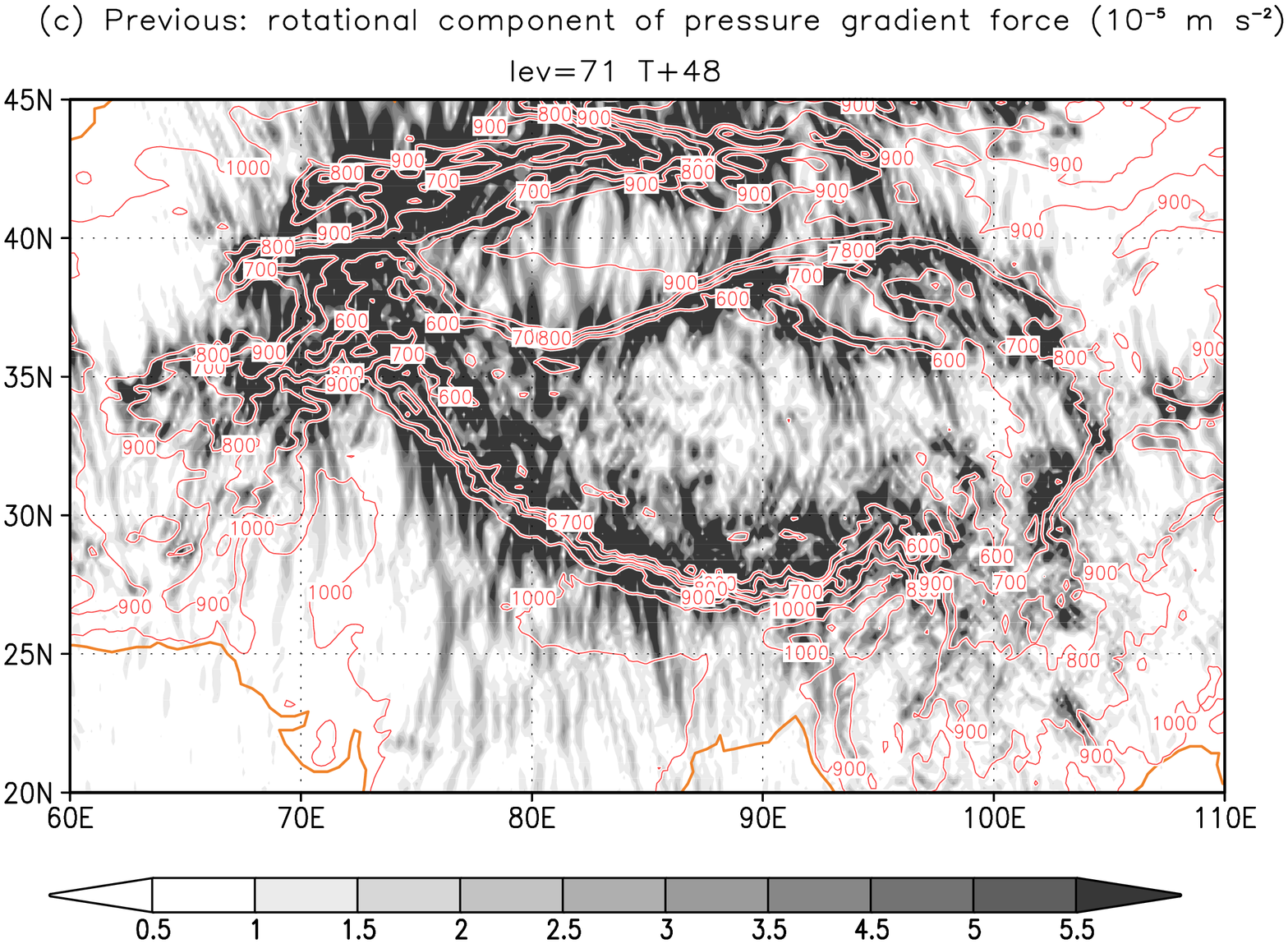}    % originally pgrad_GSM1603.eps
 \includegraphics[width=0.35\textwidth,angle=270]{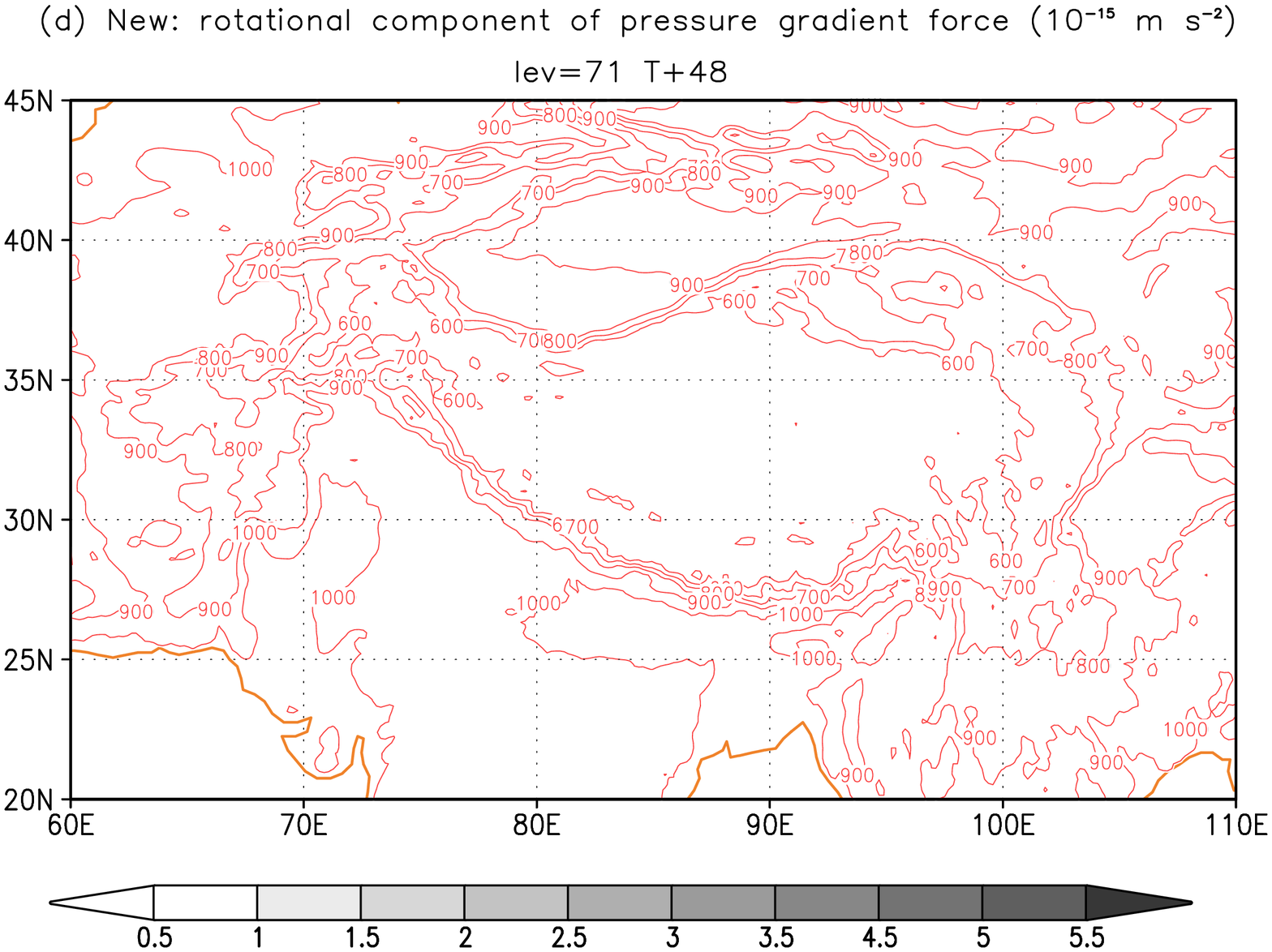}    % originally pgrad_GSM1705.eps
 \caption{
 Maps, plotted for the Himalayas region, of (a,b) vorticity (units:
 $10^{-6} \mathrm{s}^{-1}$) and (c,d) the curl of pressure gradient
 (units: $10^{-5}\mathrm{m s}^{-2}$ in (c), $10^{-15}\mathrm{m s}^{-2}$
 in (d)), both at 71st model level ($\sim$ 40 hPa), from 48-hour
 forecasts produced by JMA-GSM with (a,c) previous and (b,d) new
 pressure gradient discretisation schemes. In (a,b) the positive and
 negative contours are drawn, respectively, with solid and dashed lines,
 and the shades indicate the strength of vorticity. In (c,d) the curl of
 pressure gradient is plotted with shades, superpose on contours in
 lighter colours that show the surface pressure (units: hPa). Note that the scale of
 shades in (d) is $10^{-10}$ times smaller than in (c) to accentuate the
 small values.  Shown are the results for the same case as in Figure
 \ref{fig:spectra}.
 }
 \label{fig:map}
\end{figure}

 Compared to the de-aliasing filter approach, the proposed method has
the advantage of not introducing ad hoc correction nor tunable
parameters. Higher order grid approach can account for nonlinear
aliasing not only from pressure gradient but from any nonlinear terms;
we remark nevertheless that it does not completely eliminate small-scale
aliasing noises in rotational component of the pressure gradient (as we
verified by running GSM with quadratic Tq639 grid and then plotting a
map similar to Figure \ref{fig:map}c, not shown) although they do not
result in spectral blocking (i.e., turn-up of power spectra near the
high-wavenumber end) since the filtering effect inherent in high order
truncation prevents their accumulation over time steps. The higher order
grid approach has another advantage of being much more cost effective
than the regular linear grid truncation, particularly at very high
resolutions, while allowing more accurate representation of small-scale
variances \citep{Malardel16}.  The higher order grid approach and
the proposed method are not mutually exclusive and can be used
altogether.

The impact of adopting this alternative discretisation on spectral
blocking is significant, as we can confirm by comparing the right and
left panels of Figure \ref{fig:spectra}. The spectral blocking observed
with the previous discretisation (Equation \ref{eq:pgrad-gsm1603}) for
the rotational component at an upper level (81st out of 100 levels,
$\sim 11$ hPa; at this level, the iso-$\eta$ surface is isobaric)
completely disappears with the proposed discretisation (Figure
\ref{fig:spectra}b). The blocking remains at the lower level (51st,
$\sim 180 $ hPa) where the iso-$\eta$ surface is close to but not
completely parallel to the isobars, but to a much diminished degree
(Figure \ref{fig:spectra}d). The impact is also visible on a map as
shown in Figure \ref{fig:map}: with the previous discretisation, the
vorticity at the 71st level ($\sim 40$ hPa) plotted over the Himalayas
exhibited small-scale noisy patterns particularly along steep orography
(Figure \ref{fig:map}a); in contrast, with the proposed discretisation, the
vorticity plot is much less noisy (Figure \ref{fig:map}b). Similarly,
the rotation (curl) of the pressure gradient computed with the previous
discretisation exhibits small-scale noises (Figure \ref{fig:map}c) but
they disappear (to machine precision) by the proposed discretisation
(Figure \ref{fig:map}d).

The proposed method is very effective in reducing spectral blocking as
shown above, but this benefit comes at the expense of additional
computational and communication cost associated with one extra spectral
transform for three-dimensional variable ($\Phi$). In the case of
JMA-GSM, the increase of total execution time due to this additional
transform was found to be relatively small, partly because $\Phi$ can be
transformed together with $T_v$ so that the number of calls to MPI
routines was not increased. 

Another potential disadvantage of the proposed approach is the loss of
pre-cancellation of compensating components in $\nabla_\eta \Phi$ and
$R_dT_v\nabla_\eta\ln{p}$ exploited in the previous formulation of
JMA-GSM (Equation \ref{eq:pgrad-gsm1603}), which may result in
degradation in the model's ability to maintain geostrophic balance,
particularly in the presence of orography. This aspect is examined in
next section using idealised test cases.

\section{Idealised experiments}
\label{sec:idealised-exp} 

To assess how the proposed modification to pressure gradient
discretisation affects the model's ability to maintain balance, we
performed two idealised test cases and compared the results obtained
from the previous and proposed methods. The model we use is the dry
dynamical core of JMA-GSM. As described in the previous section, it is a
spectral SISL hydrostatic primitive equations (HPE) model discretised
with spherical-harmonics-based spectral representation in the horizontal
and finite differencing on 100 $\eta$-levels in the vertical extending
from surface up to 0.01 hPa. The time step $\Delta t$ is taken as 720 s
regardless of the horizontal resolution. Further details of the model
can be found in Section 3.1 of \cite{yukimoto11} and Section 3.2.2 of
\cite{jma-outline13}.

\subsection{Maintenance of resting atmosphere  in the presence of orography}

To highlight the impact of modifying pressure gradient discretisation
only, we first conducted a maintenance test of resting atmosphere where,
as in e.g. \cite{Klemp11}, the model is initialised with a steady state
in equilibrium with no wind. Ideally, the solution should stay at rest,
but in the model the atmosphere may start to move since the discretised
pressure gradient is not necessarily zero in the presence of orography.

\begin{figure}[htbp]
 \centering
 \includegraphics[width=0.33\textwidth,angle=270]{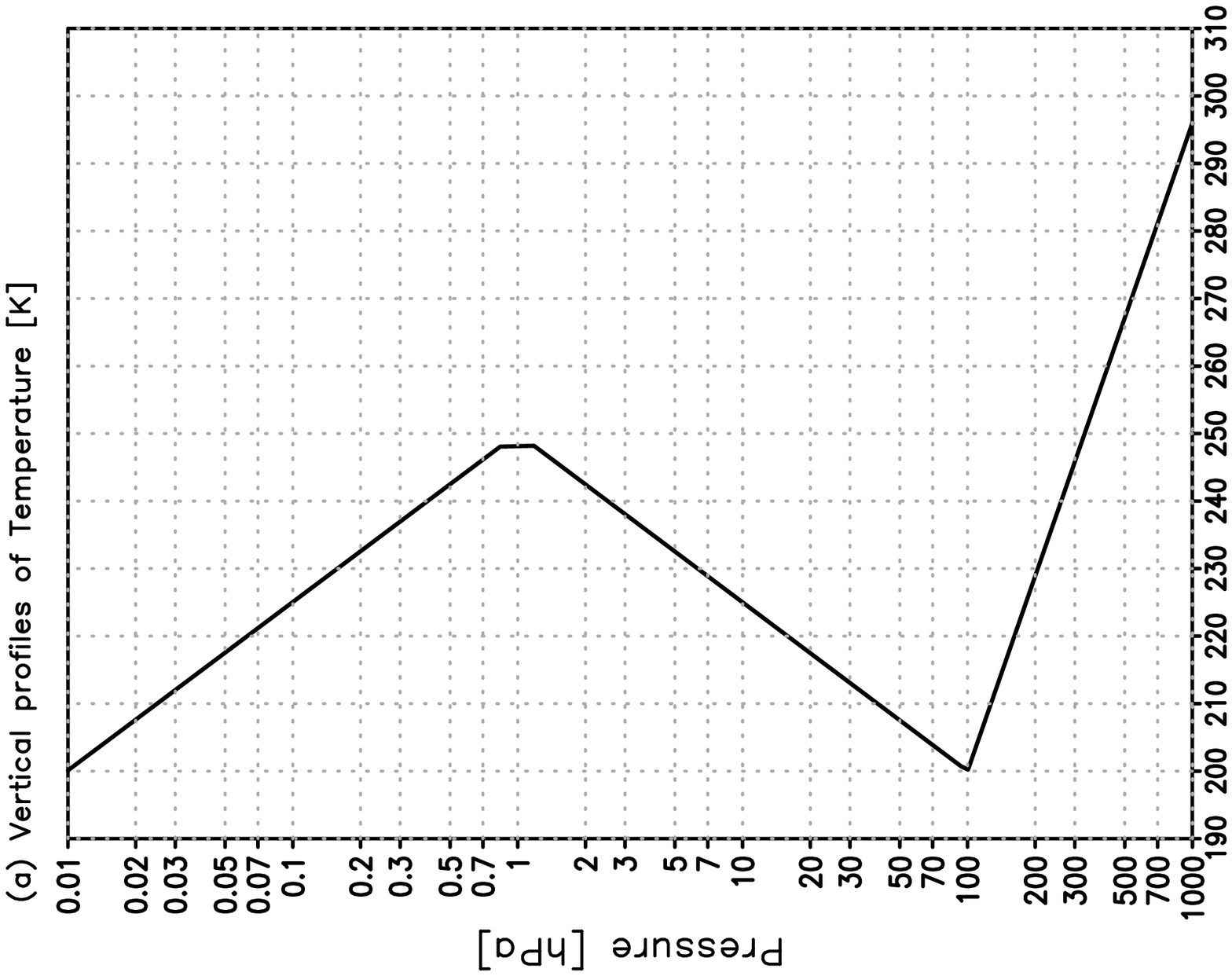} % originally tini_calm.eps
 \includegraphics[width=0.33\textwidth,angle=270]{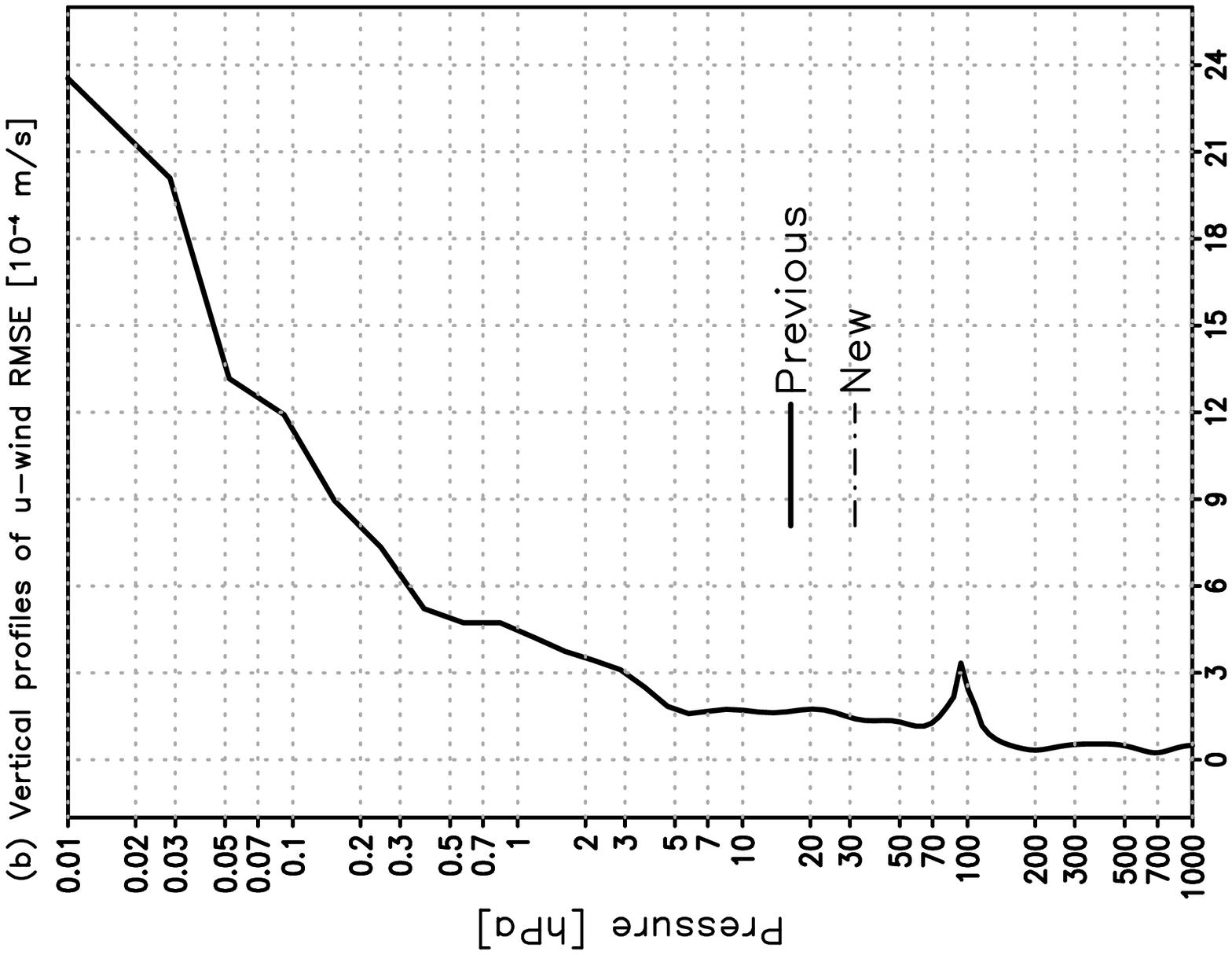} % originally uerr_calm.eps
 \includegraphics[width=0.33\textwidth,angle=270]{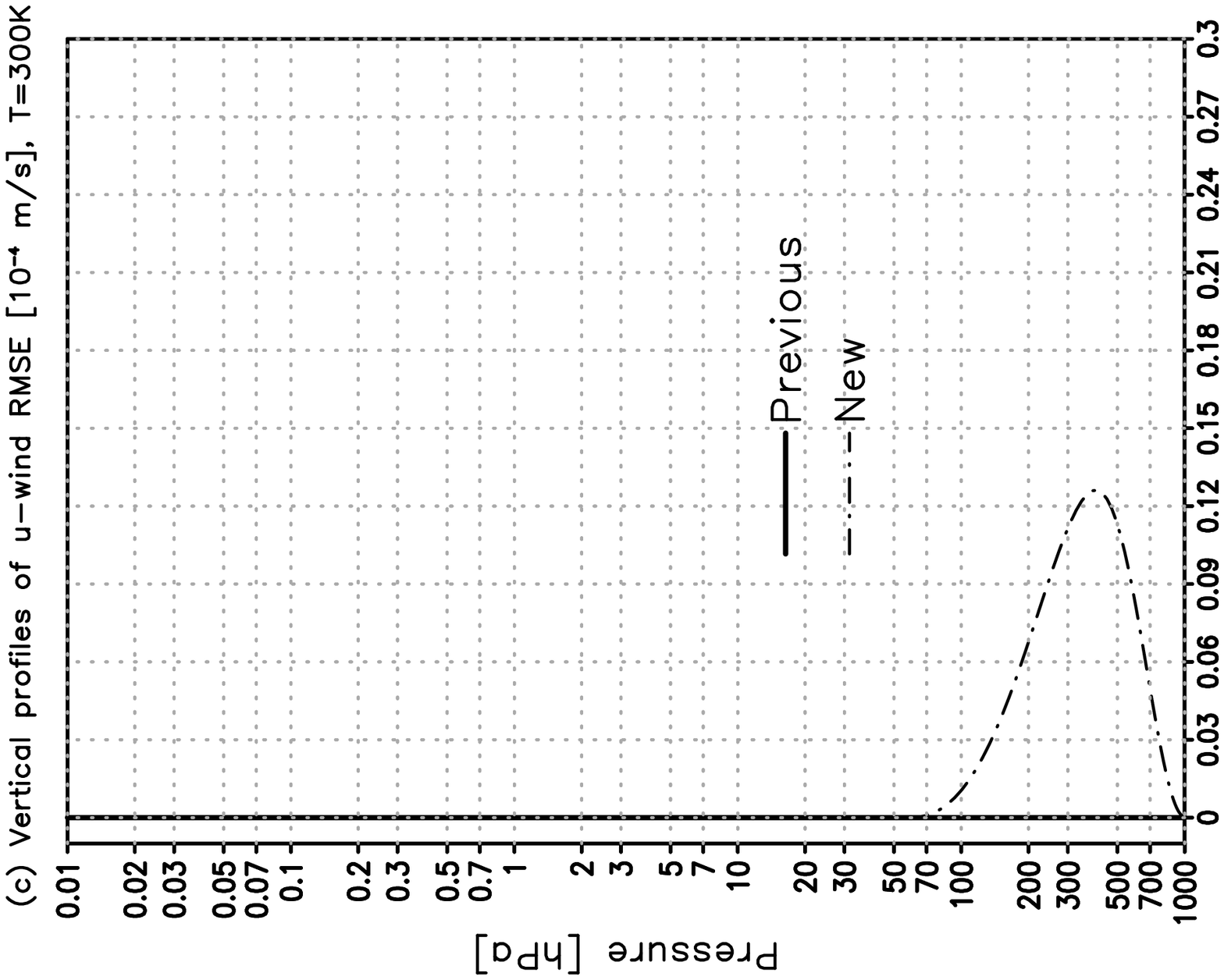} % originally uerr_calm.eps
 \caption{
 (a) Temperature profile prescribed in the resting-atmosphere
 maintenance test. (b) Profiles of the root-mean-square zonal winds
 horizontally averaged over the globe from 5-day forecasts (units:
 $10^{-4} \mathrm{m}{s}^{-1}$) produced using the previous and new pressure
 gradient discretisation schemes run at Tl319 horizontal resolution. (c)
 As in (b), but with the temperature profile replaced by a constant
 profile at 300 K. Note that, in (b), the two lines are
 indistinguishable since they collapsed onto a single line.
 }
 \label{fig:exp-rest}
\end{figure}

In this test, we prescribe a temperature profile as a function of
pressure only that mimics the real atmosphere, as shown in Figure
\ref{fig:exp-rest}a. 
% Note that, in a hydrostatic model, if the temperature
% depends only on pressure, the geopotential also depends only on pressure
% and thus the pressure gradient vanishes.
As the surface geopotential $\Phi_s$, we prescribe the two-dimensional
bell defined by Equation 9 of \cite{Wedi09} with the radius of the
Earth $a$ same as in the operation (no `small-planet' setup) and the
mountain half width $L_\lambda$ set to $0.12a$. The surface pressure
$p_s$ is determined from the primitive equations so that the pressure
gradient vanishes in the absence of winds.

The test is performed using JMA-GSM with the horizontal resolution of
Tl319. The errors, measured as the square root of the horizontal average
of the squared zonal wind from 5-day forecast, are shown for each
vertical level in Figure \ref{fig:exp-rest}b. The error from the
previous method (thick line) and that from the proposed method (thin
line) collapse onto a single profile, meaning that the model's ability
to maintain the state at rest is not harmed by using the proposed
method. While the proposed scheme is as accurate as the original scheme
for the realistic temperature profile, the original scheme did result in
much smaller errors when the temperature profile shown in Figure
\ref{fig:exp-rest}a was replaced by a contrived iso-thermal temperature
profile at 300 K (Figure \ref{fig:exp-rest}c). This, together with the
results for a more realistic profile, indicate that the pre-cancellation
of compensating components exploited in the previous scheme is only
effective for idealised (unrealistic) situations.

\subsection{Jablonowski-Williamson steady sate test}

To further assess the impact of using the proposed pressured gradient
discretisation, we then conducted the steady-state test proposed by
\cite{JW06} which we believe is more holistic in that not only pressure
gradient terms but also other terms of the governing equations play a
role. In this test case the model is initialised with an analytically
defined steady-state solution and integrated up to 9 days, to examine to
what extent the model is able to maintain this steady state. The steady
state is baroclinically unstable, so that imbalances induced by
discretisation error will amplify and become detectable.

The steady-state test is performed using JMA-GSM with three different
horizontal resolutions of Tl63, Tl319 and Tl959 (which correspond to
grid spacing of $\sim 300$ km, $\sim 60$ km and $\sim 20$ km at the
Equator, respectively), all with the operational 100 vertical
levels. The globally-averaged $l_2$-deviation of the zonal wind field
$u$ from their zonal mean (Equation 14 of \cite{JW06}: the error
associated with growth of baroclinic waves) was larger with the proposed
method than with the previous method but only by about 20--30\% for
Tl319 and Tl959 resolutions (at Tl63 resolution, the errors were almost
identical; figures omitted). The temporal degradation of the zonal mean
zonal wind field (Equation 15 of \cite{JW06}: the error associated with
geostrophic adjustment in the discrete system) quickly saturated by day
1 and stayed around $2.5\times 10^{-3} \mathrm{m}\mathrm{s}^{-1}$ for
any resolution with both the proposed and previous methods (figures
omitted; the error curves were very similar to the one shown in Figure 5
of \cite{HU18}).

The results from the two sets of idealised experiments all suggest that
the new pressure gradient discretisation is as accurate as the previous
one except in the contrived isothermal setup, motivating us to conduct
full NWP experiments at quasi-operational setup that we describe below.

\section{Cycled NWP experiment}
\label{sec:nwp-exp}

The impact of the proposed pressure gradient discretisation upon
forecast performance was assessed by conducting cycled NWP experiments
following the standard practice at JMA. We conducted cycled NWP
experiments for two distinct periods (SUMMER and WINTER) that each
covers more than a month.
The SUMMER experiment consists of 6-hourly data assimilation cycle that
begins on July 10, 2015 at 00 UTC and ends on September 11, 2015 at 18
UTC, and extended forecasts up to 11 days that are launched daily only
at 12 UTC from July 21, 2015 till August 31, 2015.  Similarly, the
WINTER experiment consists of data assimilation cycle that begins on
December 10, 2014 at 00UTC and ends on February 11, 2015 at 18 UTC, and
extended forecasts launched daily at 12 UTC from December 21, 2014 to
January 31, 2015.

The update of pressure gradient discretisation only resulted in neutral
impact in terms of all headline scores (figures omitted). The
first-guess fit to observations (O-B departures) were also closely
examined, with particular focus on stratosphere-sensitive instruments
like radiances from AMSU-A and microwave sounders and GNSS
radio-occultation, but no significant differences were detected (figures
omitted). From these results we conclude that the new discretisation
successfully reduces spectral blocking in vorticity without harming
forecast performance.

\section{Conclusions}
\label{sec:conclusions}

The vertical discretisation of the pressure gradient terms adopted by
JMA-GSM following \cite{SB81}, which is typical of  many global spectral HPE
models on $\eta$ hybrid coordinate, is found to incur spectral blocking
in the rotational winds, particularly in the upper layers where the
iso-$\eta$ levels are (close to) isobaric. This results from the
inability of grid-space evaluation of geopotential gradient $\nabla_\eta
\Phi$ to preserve its rotation-free property, which is easily fixed by
spectral evaluation of the $\nabla_\eta \Phi$ term as we proposed in
section \ref{sec:remedy}. The proposed method was tested in both
idealised and quasi-operational experiments and was found to
significantly reduce the spectral blocking without harming forecast
performance. The proposed scheme was incorporated into JMA's operational
deterministic forecasting system in its May 2017 update
\citep{Yonehara18}. The presence of spectral blocking in spectral SISL
models on linear grid, and the importance of assuring rotation-free
property on the discretised gradient operator, have both been
well-recognised, but the connection between the two appears not to be
well clarified in the literature. Since the pressure gradient
discretisation that we described in Section \ref{sec:SB81} is quite
generic, the solution that we described in Section \ref{sec:remedy} should
be widely applicable to other spectral SISL models as well.

The finding documented in this note highlights the importance of
preserving rotation-free property in the discretised gradient operators.
This is not a new discovery, and has already been emphasised by
\cite{StaniforthThuburn12} as one of the essential desiderata for future
dynamical cores.  This is a challenging task, especially for grid-based
discretisation, but, promising progress has already been made, for
example, by \cite{Thuburn14} and \cite{Weller14}. 

An interesting and useful lesson that we draw from the presented results
is that pressure gradient discretisation as a whole can bear rotational
component even if the mimetic ``rotation-free gradient'' property
$(\nabla\times\nabla=0)$ is satisfied operator-wise (as with spectral
representation). Given the current trend of high-performance computing
where growth of computing capacity relies increasingly on massive
parallelism, spectral models are predicted to face serious scalability
issue, and many centers, including JMA, are exploring transition to (or
have already transitioned to) a grid-based model with better data
locality \citep[e.g.,][]{HU18,FVM1.0}. The lesson that we learned here will serve
nicely as a guiding principle in deciding which discretisation to use
out of many possibilities.

\section*{ACKNOWLEDGEMENTS}
The authors thank colleagues at Numerical Prediction Division of JMA
%, most notably Dr. Hitoshi Yonehara, Misters Takashi Kadowaki, Takafumi
%Kanehama (currently detailed to ECMWF) and Yukihiro Kuroki 
for helpful discussion and support; the authors also wish to thank
Dr. Nils Wedi of ECMWF for his invigorating comments on an early version
of the manuscript.

\section*{CONFLICT OF INTEREST}

Authors declare no conflict of interest.

%\printendnotes

% Submissions are not required to reflect the precise reference formatting of the journal (use of italics, bold etc.), however it is important that all key elements of each reference are included.
%\bibliography{pgrad}

\clearpage
%%%%%%%%%% Figures %%%%%%%%%
%\clearpage
\section*{Figures}
%\clearpage
\clearpage

%%%%%%%%%%%%%%%%%%

\end{document}